\pdfoutput=1
\documentclass[a4paper]{article}
\usepackage{INTERSPEECH2019}
\usepackage{subfigure}
\usepackage{array,multirow}
\usepackage{multirow,bigdelim}
\usepackage{bm}

\usepackage{amsmath}
\usepackage{hyperref}
\input{def.set}

\title{Cascaded Cross-Module Residual Learning \\\ towards Lightweight End-to-End Speech Coding}    
\name{Kai Zhen$^{1,2}$, Jongmo Sung$^3$, Mi Suk Lee$^3$, Seungkwon Beack$^3$, Minje Kim$^{1,2}$}
\address{$^1$Indiana University, School of Informatics, Computing, and Engineering, Bloomington, IN \newline
$^2$Indiana University, Cognitive Science Program, Bloomington, IN \newline
$^3$Electronics and Telecommunications Research Institute, Daejeon, South Korea\newline}
\email{zhenk@iu.edu, jmseong@etri.re.kr, lms@etri.re.kr, skbeack@etri.re.kr, minje@indiana.edu}

\begin{document}
\maketitle
\begin{abstract}
Speech codecs learn compact representations of
speech signals to facilitate data transmission. Many recent deep neural network (DNN) based end-to-end speech codecs achieve low bitrates and high perceptual
quality at the cost of model complexity. We propose
a cross-module residual learning (CMRL) pipeline as a module carrier with each module
reconstructing the residual from its preceding modules.
CMRL differs from other DNN-based speech codecs, in that rather than modeling speech compression problem in
a single large neural network, it optimizes a series of less-complicated modules in a two-phase training scheme.
The proposed method shows better objective performance than AMR-WB and the state-of-the-art DNN-based speech codec with a similar network architecture. As an end-to-end model, it takes raw PCM signals as an input, but is also compatible with linear predictive coding (LPC), showing better subjective quality at high bitrates than AMR-WB and OPUS. The gain is achieved by using only 0.9 million trainable parameters, a significantly less complex architecture than the other DNN-based codecs in the literature. 
\end{abstract}

\noindent\textbf{Index Terms}: speech coding, deep neural network, entropy coding, residual learning

\section{Introduction}

\label{sec:intro}
Speech coding, where the encoder converts the  speech signal into bitstreams and the decoder synthesizes reconstructed signal from received bitstreams, serves an important role for various purposes: to secure a voice communication \cite{classic_book_audiocoding_mpeg}\cite{classic_algo_audiocoding_mpeg1}, to facilitate data transmission \cite{classic_book_audiocoding}, etc. There have been various conventional speech coding methodologies, including linear predictive coding (LPC) \cite{lpc}, adaptive encoding \cite{atal1970adaptive}, and perceptual weighting \cite{schroeder1985code} among other domain specific knowledge about the speech signals, that are used to construct classic codecs, such as AMR-WB \cite{amrwb} and OPUS \cite{opus} with high perceptual quality.

Since the last decade, 
data-driven approaches have vitalized the use of deep neural networks (DNN) for speech coding.
A speech coding system can be formulated by DNN as an autoencoder (AE) with a code layer discretized by vector quantization (VQ) \cite{makhoul1985vector} or bitwise network techniques \cite{KimMJ2015icmlw}, etc. 
Many DNN methods \cite{deng2010binary}\cite{new_paper_nn_lowbitrate_2016} take inputs in time-frequency (T-F) domain from short time Fourier transform (STFT) or modified discrete cosine transform (MDCT), etc.
Recent DNN-based codecs \cite{new_paper_google_wavenet_2017}\cite{new_paper_bloombergEndtoEnd}\cite{liu2018wavenet}\cite{google_wavenet_2019} model speech signals in time domain directly without T-F transformation. They are referred to as end-to-end methods, yielding competitive performance comparing with current speech coding standards, such as AMR-WB \cite{amrwb}.

While DNN serves a powerful parameter estimation paradigm, they are computationally expensive to run on smart devices. Many DNN-based codecs achieve both low bitrates and high perceptual quality, two main targets for speech codecs \cite{salami1997itu}\cite{recommendation2003722}\cite{neuendorf2012mpeg}, but with a high model complexity. 
A WaveNet based variational autoencoder (VAE) \cite{google_wavenet_2019} outperforms other low bitrate codecs in the listening test, however, with 20 millions parameters, a too big model for real-time processing in a resource-constrained device. Similarly, codecs built on SampleRNN \cite{samplernn}\cite{rnn_codec_2019} can also be energy-intensive.   

Motivated by DNN based end-to-end codecs \cite{new_paper_bloombergEndtoEnd} and residual cascading \cite{johnston2018improved}\cite{schuller2001lossless}, this paper proposes a ``cross-module" residual learning (CMRL) pipeline, which can lower the model complexity while maintaining a high perceptual quality and compression ratio. CMRL hosts a list of less-complicated end-to-end speech coding modules. Each module learns to recover what is failed to be reconstructed by its preceding modules. CMRL differs from other residual learning networks, e.g. ResNet \cite{he2016deep}, in that rather than adding identical shortcuts between layers, CMRL cascades residuals across a series of DNN modules. We introduce a two-round model training scheme to train CMRL models. In addition, we also show that CMRL is compatible with LPC by having it as one of the modules. 
With LPC coefficients being predicted, CMRL recovers the LPC residuals which, along with the LPC coefficients, synthesize the decoded speech signal at the receiver side.

The evaluation of the propose method is threefold: objective measures, subjective assessment and model complexity. Comparing with AMR-WB, OPUS, and the recently proposed end-to-end system \cite{new_paper_bloombergEndtoEnd}, CMRL showed promising performance both in objective and subjective quality assessments. As for complexity, CMRL contains only 0.9 million model parameters, significantly less complicated than the WaveNet based speech codec \cite{google_wavenet_2019} and the end-to-end baseline \cite{new_paper_bloombergEndtoEnd}. 


\begin{figure*}[t]
    \centering
    \includegraphics[width=.92\textwidth]{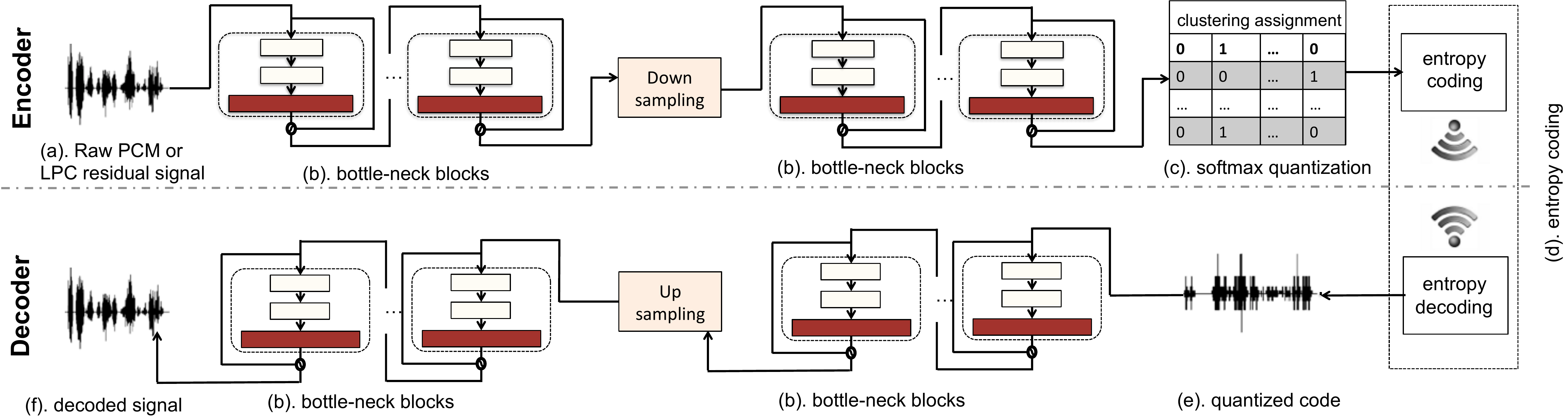}
    \caption{A schematic diagram for the end-to-end speech coding component module: some channel change steps are omitted.}
    \label{fig:end2end}
\end{figure*}

\section{Model description}
Before introducing CMRL as a module carrier, we describe the component module to be hosted by CMRL.

\subsection{The component module}
Recently, an end-to-end DNN speech codec (referred to as Kankanahalli-Net) has shown competitive performance comparable to one of the standards (AMR-WB) \cite{new_paper_bloombergEndtoEnd}. We describe our component model derived from Kankanahalli-Net that consists of bottleneck residual learning \cite{he2016deep}, soft-to-hard quantization \cite{agustsson2017soft}, and sub-pixel convolutional neural networks for upsampling \cite{shi2016real}. Figure \ref{fig:end2end} depicts the component module.

\subsubsection{Four non-linear mapping types}
In the end-to-end speech codec, we take $S=512$ time domain samples per frame, 32 of which are windowed by the either left or right half of a Hann window and then overlapped with the adjacent ones. This forms the input to the first 1-D convolutional layer of $C$ kernels, whose output is a tensor of size $S\times C$ . 



\begin{figure}[t]
\includegraphics[width=.46\textwidth]{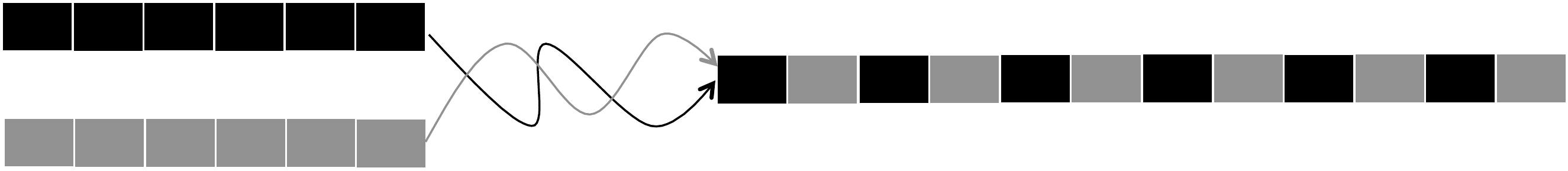}
  \caption{The interlacing-based upsampling process.} 
  \label{fig:resnet}
\end{figure}

There are four types of non-linear transformations involved in this fully convolutional network: downsampling, upsampling, channel changing, and residual learning. The downsampling operation reduces $S$ down to $S/2$ by  setting the stride $d$ of the convolutional layer to be $2$, which turns an input example $S\times C$ into $S/2 \times C$. The original dimension $S$ is recovered in the decoder with recently proposed sub-pixel convolution \cite{agustsson2017soft}, which forms the upsampling operation. The super-pixel convolution is done by interlacing multiple feature maps to expand the size of the window (Figure ~\ref{fig:resnet}). In our case, we interlace a pair of feature maps, and that is why in Table \ref{tab:topo} the upsampling layer reduces the channels from 100 to 50 while recovers the original 512 dimensions from 256.

In this work, to simplify the model architecture we have identical shortcuts only for cross-layer residual learning, while Kankanahalli-Net employs them more frequently. Furthermore, inspired by recent work in source separation with dilated convolutional neural network \cite{tan2018gated}, we use a ``bottleneck" residual learning block to further reduce the number of parameters. This can lower the amount of parameters, because the reduced number of channels within the bottleneck residual learning block decreases the depth of the kernels. See Table \ref{tab:topo} for the size of our kernels. Likewise, the input $S \times 1$ tensor is firstly converted to a $S \times C$ feature map, and then downsampled to $S/2 \times C$. Eventually, the code vector shrinks down to $S/2 \times 1$. The decoding process recovers it back to a signal of size $S \times 1$, reversely.




\subsubsection{Softmax quantization:}
The coded output from each encoder is still a real-valued vector of size $S/2$. Softmax quantization \cite{agustsson2017soft} performs scalar quantization by assigning each real value to the nearest representative (Figure ~\ref{fig:end2end} (c)). In the proposed system, softmax quantization maps the input scalar to one of the 32 clusters, or quantization levels, which requires $\log_2 32 = 5$ bits per dimension. Huffman coding further reduces the bitrate \cite{huffman1952method}. 

\begin{figure*}
    \centering
    \includegraphics[width=0.9\textwidth]{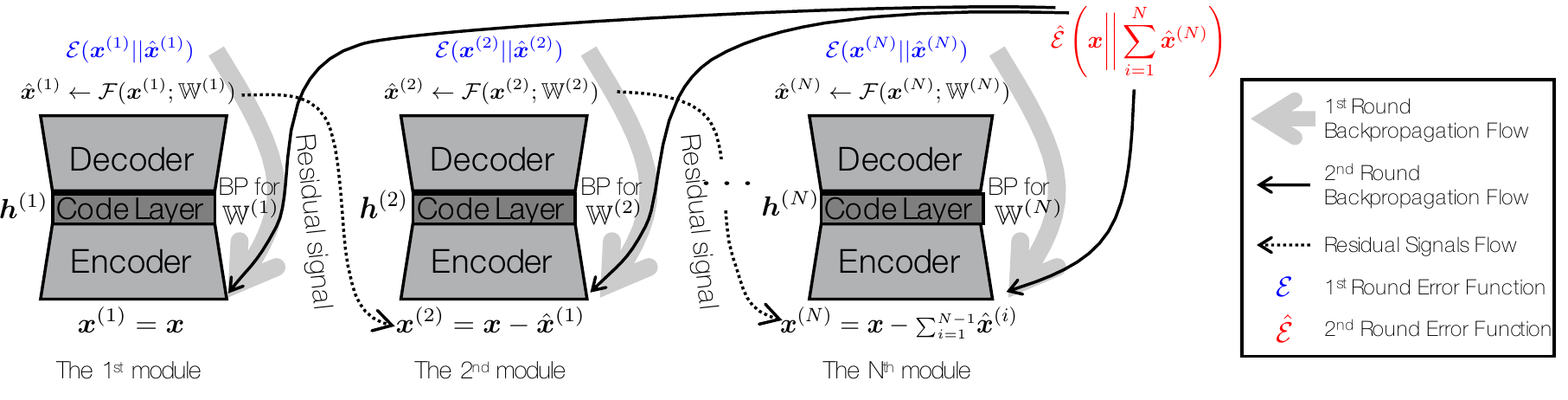}
    \vspace{-0.1in}
    \caption{Cross-module residual learning pipeline}
    \label{fig:themodel}
\end{figure*}

\subsection{The module carrier: CMRL}
Figure  \ref{fig:themodel} shows the proposed cascaded cross-module residual learning (CMRL) process. In CMRL, each module does its best to reconstruct its input. The procedure in the $i$-th module is denoted as $\calF(\bx^{(i)}; \mathbb{W}^{(i)})$, which estimates the input as $\hat{\bx}^{(i)}$. The input for the $i$-th module is defined as 
\begin{equation}\label{eq:input}
\bx^{(i)} = \bx - \sum\limits_{j=1}^{i-1}\hat{\bx}^{(j)},
\end{equation}
where the first module takes the input speech signal, i.e., $\bx^{(1)}=\bx$. The meaning is that each module learns to reconstruct the residual which is not recovered by its preceding modules. Note that module homogeneity is not required for CMRL: for example, the first module can be very shallow to just estimate the envelope of MDCT spectral structure while the following modules may need more parameters to estimate the residuals.

Each AE decomposes into the encoder and decoder parts:
\begin{align}
    \bh^{(i)}=\calF_\text{enc}(\bx^{(i)}; \mathbb{W}^{(i)}_\text{enc}), \quad    \hat{\bx}^{(i)}=\calF_\text{dec}(\bh^{(i)}; \mathbb{W}^{(i)}_\text{dec}),
\end{align}
where $\bh^{(i)}$ denotes the part of code generated by the $i$-th encoder, and $\mathbb{W}^{(i)}_\text{enc} \cup \mathbb{W}^{(i)}_\text{dec} = \mathbb{W}^{(i)}$.

{\em\textbf{The encoding process}}: For a given input signal $\bx$, the encoding process runs all $N$ AE modules in a sequential order. Then, the bitstring is generated by taking the encoder outputs and concatenating them:  $\bh=\Big[{\bh^{(1)}}^\top,{\bh^{(2)}}^\top,\cdots,{\bh^{(N)}}^\top\Big]^\top$. 

{\em\textbf{The decoding process}}: Once the bitstring is available on the receiver side, all the decoder parts of the modules, $\calF_\text{dec}(\bx^{(i)}; \mathbb{W}^{(i)}_\text{dec}) ~ \forall N$, run to produce the reconstructions which are added up to approximate the initial input signal with the global error defined as
\begin{equation}\label{eq:totalErr}
\hat{\calE}\left(\bx\bigg|\bigg| 
\sum_{i=1}^{N}\hat{\bx}^{(i)}
\right).
\end{equation}

\subsubsection{The two-round training scheme}
{\em\textbf{Intra-module greedy training:}}
 We provide a two-round training scheme to make CMRL optimization tractable.  The first round adopts a {\em greedy} training scheme, where each AE tries its best to minimize the error: $\argmin\limits_{\mathbb{W}^{(i)}} \calE(\bx^{(i)}||\calF(\bx^{(i)}; \mathbb{W}^{(i)}))$. The greedy training scheme echoes a divide-and-conquer manner, leading to an easier optimization for each module. The thick gray arrows in Figure  \ref{fig:themodel} show the flow of the backpropagation error to minimize the individual module error with respect to the module-specific parameter set $\mathbb{W}^{(i)}$. 


{\em\textbf{Cross-module finetuning:}}
The greedy training scheme accumulates module-specific error, which the earlier modules do not have a chance to reduce, thus leading to a suboptimal result. Hence, the second-round cross-module finetuning follows to further improve the performance by reducing the total error:
\begin{equation}\label{eq:2round}
\argmin\limits_{\mathbb{W}^{(1)}\cdots\mathbb{W}^{(N)}} \hat{\calE}\left(\bx\bigg|\bigg|\sum\limits_{i=1}^N\calF\big(\bx^{(i)}; \mathbb{W}^{(i)}\big)\right).
\end{equation}
During the finetuing step, we first (a) initialize the parameters of each module with those estimated from the greedy training step (b) perform cascaded feedforward on all the modules sequentially to calculate the total estimation error in \eqref{eq:totalErr} (c) backpropagate the error to update parameters in all modules altogether (thin black arrows in Figure  \ref{fig:themodel}). Aside from the total reconstruction error \eqref{eq:totalErr}, we inherit Kankanahalli-Net's other regularization terms, i.e., perceptual loss, quantization penalty, and entropy regularizer. 

\subsection{Bitrate and entropy coding}
The bitrate is calculated from the concatenated bitstrings from all modules in CMRL. Each encoder module produces $S/d$ quantized symbols from the softmax quantization process (Figure \ref{fig:end2end} (e)), where the stride size $d$ divides the input dimensionality. Let $c^{(i)}$ be the average bit length per symbol after Huffman coding in the $i$-th module. Then, $c^{(i)}S/d$ stands for the bits per frame. By dividing the frame rate, $(S-o)/f$, where $o$ and $f$ denote the overlap size in samples and the sampling rate, respectively, the bitrates per module add up to the total bitrate: $\xi_\text{LPC} + \sum_{i=1}^{N}\frac{fcS}{(S-o)d}$, 
where the overhead to transmit LPC coefficients is $\xi_\text{lpc}$=2.4kbps, which is $0$ for the case with raw PCM signals as the input.

By having the entropy control scheme proposed in Kankanahalli-Net as the baseline to keep a specific bitrate, we further enhance the coding efficiency by employing the Huffman coding scheme on the vectors.
Aside from encoding each symbol (i.e., the softmax result) separately, encoding short sequences can further leverage the temporal correlation in the series of quantized symbols, especially when the entropy is already low \cite{witten1987arithmetic} \cite{welch1986error}. We found that encoding a short symbol sequence of adjacent symbols, i.e., two symbols, can lower down the average bit length further in the low bitrates.

\begin{table}[h]
\centering
\scriptsize
\caption{Architecture of the component module as in Figure ~\ref{fig:end2end}. Input and output tensors sizes are represented by (width, channel), while the kernel shape is (width, in channel, out channel).}
\setlength\tabcolsep{5.6pt}
\begin{tabular}{ c|c|c|c }
 \hline
 Layer &Input shape & Kernel shape & Output shape\\
 \hline
Change channel & (512, 1) & (9, 1, 100) &(512, 100) \\
\hline
1st bottleneck & (512, 100) & \begin{tabular}{cc}\rule[6pt]{0pt}{0pt}(9, 100, 20)  &\rdelim]{3}{5mm}[$\times$2]\\ (9, 20, 20)  &  \\(9, 20, 100) &\rule[-2pt]{0pt}{0pt}\end{tabular} &(512, 100)  \\\hline
Downsampling & (512, 100) & (9, 100, 100) &(256, 100) \\
\hline
2nd bottleneck & (256, 100) & \begin{tabular}{cc}\rule[6pt]{0pt}{0pt}(9, 100, 20)  &\rdelim]{3}{5mm}[$\times$2]\\ (9, 20, 20)  &  \\(9, 20, 100) &\rule[-2pt]{0pt}{0pt}\end{tabular} &(256, 100)  \\
\hline
Change channel & (256, 100) & (9, 100, 1) &(256, 1) \\
\hline
\hline
Change channel & (256, 1) & (9, 1, 100) &(256, 100) \\
\hline
1st bottleneck & (256, 100) & \begin{tabular}{cc}\rule[6pt]{0pt}{0pt}(9, 100, 20)  &\rdelim]{3}{5mm}[$\times$2]\\ (9, 20, 20)  &  \\(9, 20, 100) &\rule[-2pt]{0pt}{0pt}\end{tabular} &(256, 100)  \\\hline
Upsampling & (256, 100) & (9, 100, 100) &(512, 50) \\
\hline
2nd bottleneck & (512, 50) & \begin{tabular}{cc}\rule[6pt]{0pt}{0pt}(9, 50, 20)  &\rdelim]{3}{5mm}[$\times$2]\\ (9, 20, 20)  &  \\(9, 20, 50) &\rule[-2pt]{0pt}{0pt}\end{tabular} &(512, 50)  \\\hline
Change channel & (512, 50) & (9, 50, 1) &(512, 1) \\
\hline
\end{tabular}
\vspace{-0.05in}
\label{tab:topo}
\end{table}

\begin{figure*}[t]
\subfigure[8.85kbps]{\includegraphics[height=1.56in]{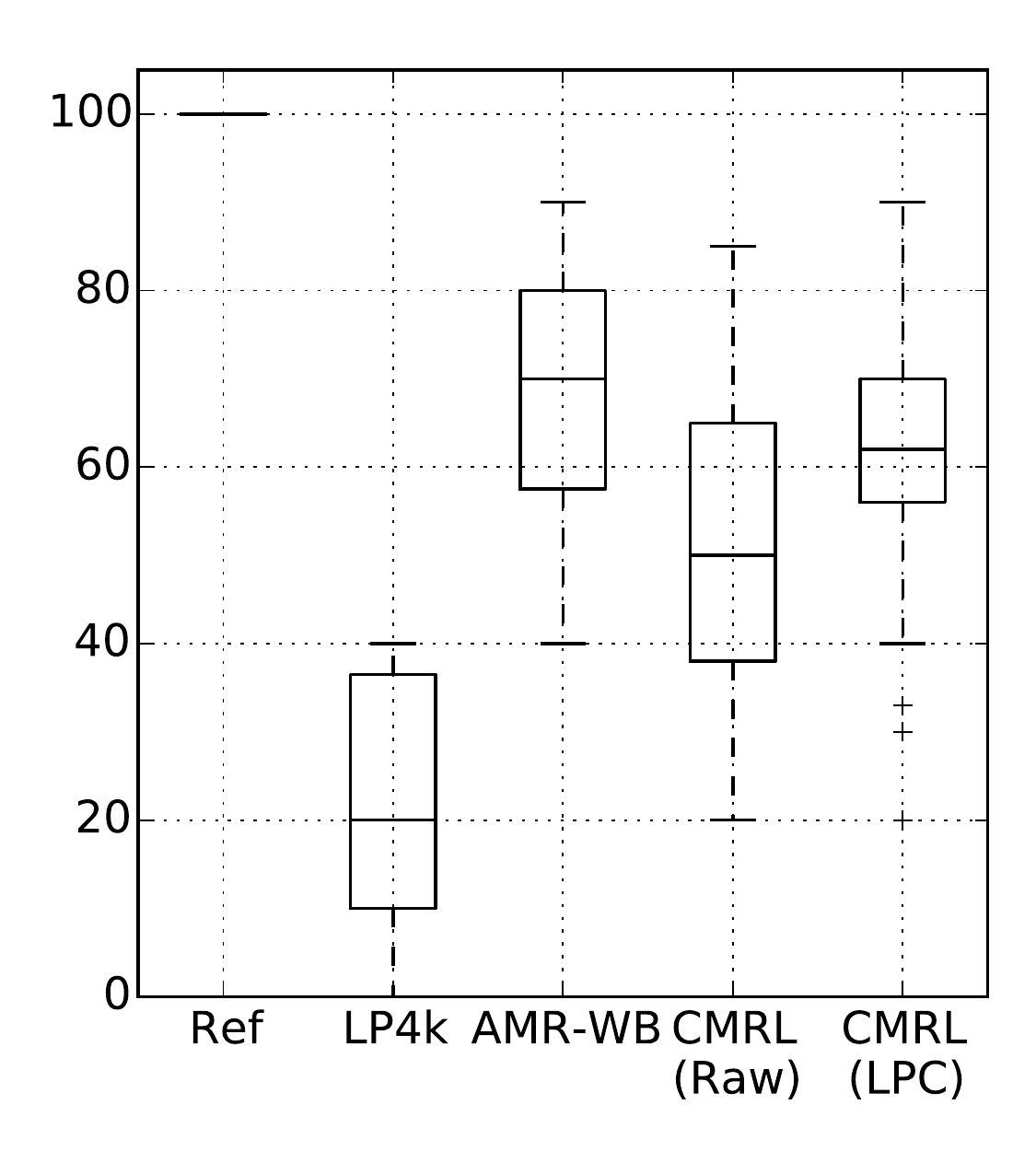}}\hspace{0.0in}
\subfigure[15.85kbps]{\includegraphics[height=1.56in]{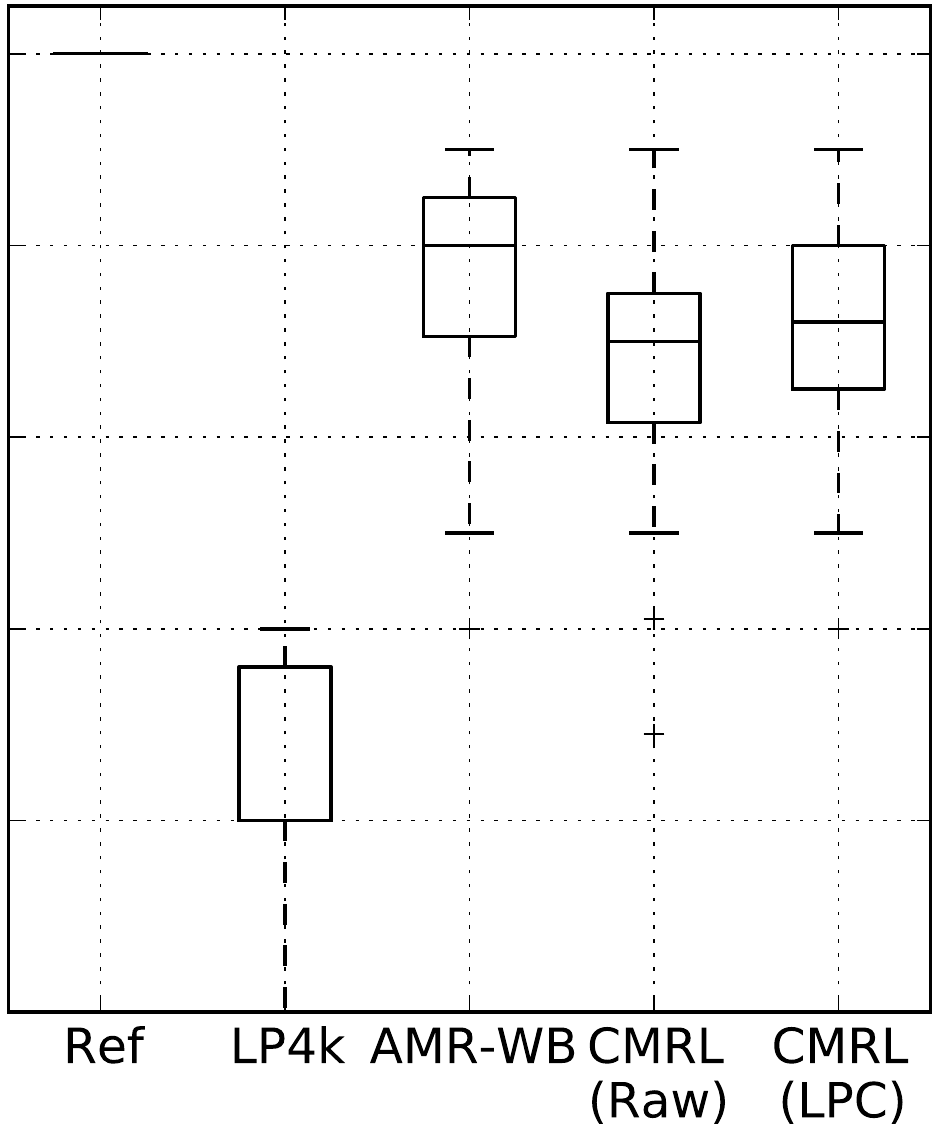}}\hspace{0.0in}
\subfigure[19.85kbps]{\includegraphics[height=1.56in]{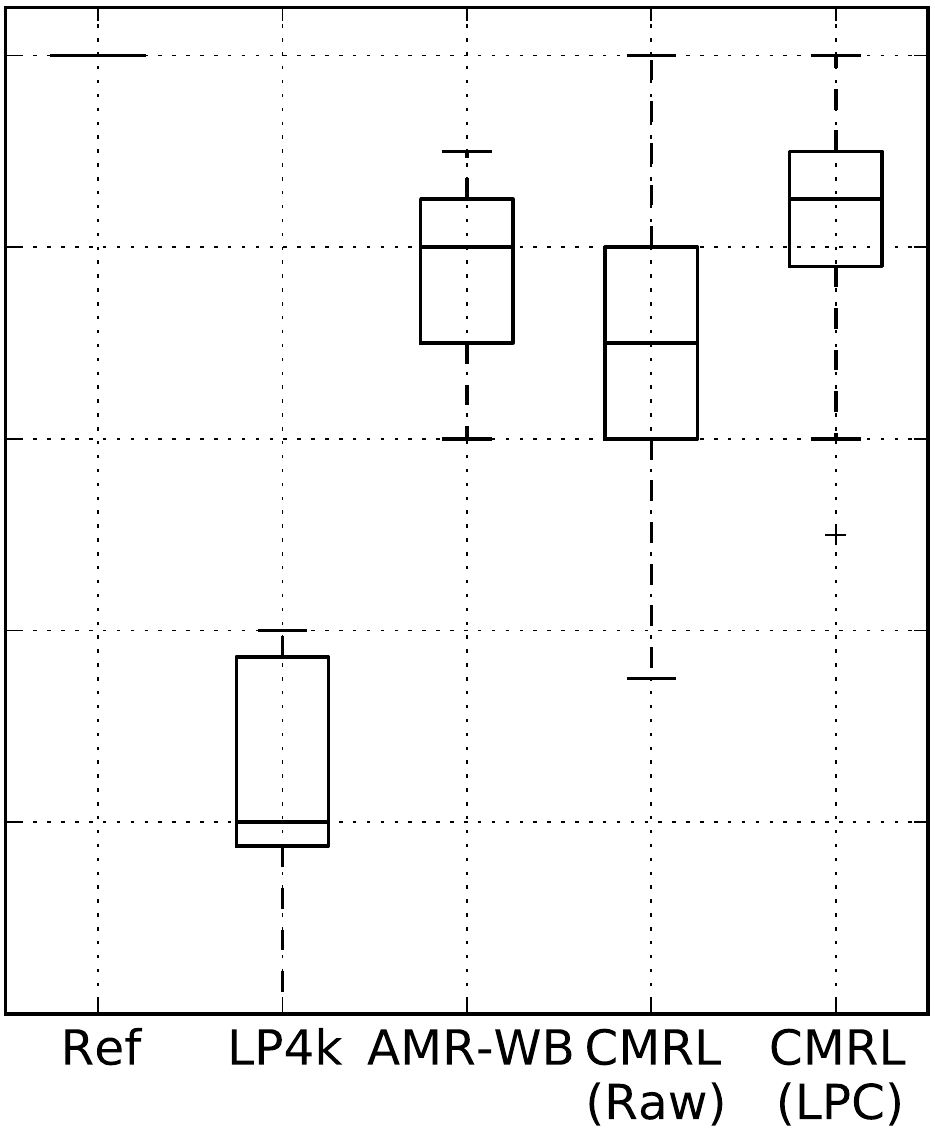}}\hspace{0.0in}
\subfigure[23.85kbps]{\includegraphics[height=1.56in]{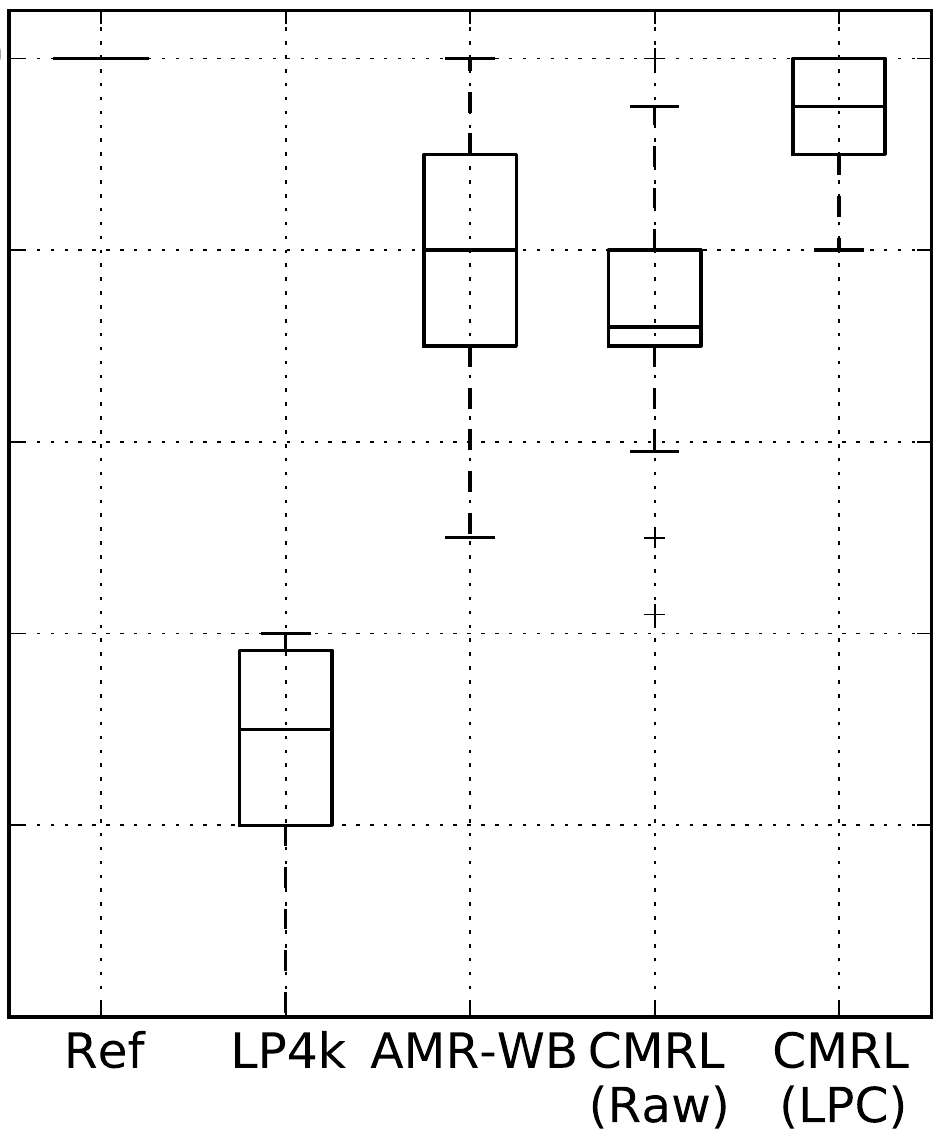}}\hspace{0.0in}
\subfigure[23.85kbps]{\includegraphics[height=1.56in]{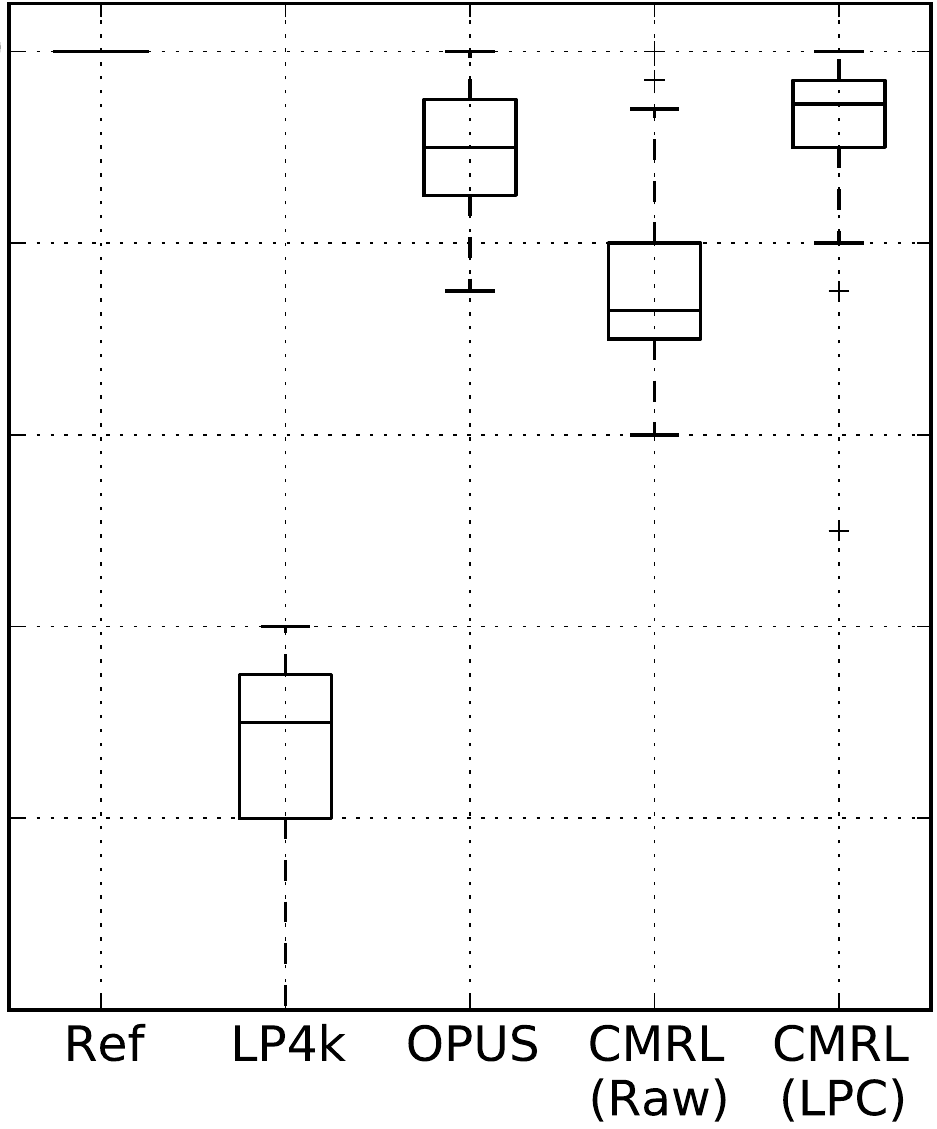}}
\vspace{-0.06in}
    \caption{MUSHRA test results. From (a) to (d): the performance of CMRL on raw and LPC residual input signals compared against AMR-WB at different bitrates. (e) An additional test shows that the performance of CMRL with the LPC input competes with OPUS, which is known to outperform AMR-WB in 23.85kbps.  }
    \label{fig:mos}
\end{figure*}

\section{Experiments}

We first show that for the raw PCM input CMRL outperforms AMR-WB and Kankanahalli-Net in terms of objective metrics in the experimental setup proposed in \cite{new_paper_bloombergEndtoEnd}, where the use of LPC was not tested. Therefore, for the subjective quality, we perform MUSHRA tests \cite{itu20031534} to show that CMRL with an LPC residual input works better than AMR-WB and OPUS at high bitrates. 

\subsection{Experimental setup}
300 and 50 speakers are randomly selected from TIMIT \cite{timit} training and test datasets, respectively. We consider two types of inputs in time-domain: raw PCM and LPC residuals. For the raw PCM input, the data is normalized to have a unit variance, and then directly fed to the model. For the LPC residual input, we conduct a spectral envelope estimation on the raw signals to get LPC residuals and corresponding coefficients. The LPC residuals are modeled by the proposed end-to-end CMRL pipeline, while the LPC coefficients are quantized and sent directly to the receiver side at 2.4 kbps. The decoding process recovers the speech signal based on the LPC synthesis procedure using the LPC coefficients and the decoded residual signals.

We consider four bitrate cases: 8.85 kbps, 15.85 kbps, 19.85 kbps and 23.85 kbps. All convolutional layers in CMRL use 1-D kernel with the size of 9 and the Leaky Relu activation. CMRL hosts two modules: each module is with the topology as in Table \ref{tab:topo}. Each residual learning block contains two bottleneck structures with the dilation rate of 1 and 2. Note that for the lowest bitrate case, the second encoder downsamples each window to 128 symbols. The learning rate is 0.0001 to train the first module, and 0.00002 for the second module. Finetuning uses 0.00002 as the learning rate, too. Each window contains 512 samples with the overlap size of 32. We use Adam optimizer \cite{kingma2014adam} with the batch size of 128 frames. Each module is trained for 30 epochs followed by finetuning until the entropy is within the target range.

\begin{figure}
   \centering
   \subfigure[]{\includegraphics[width=.77\columnwidth]{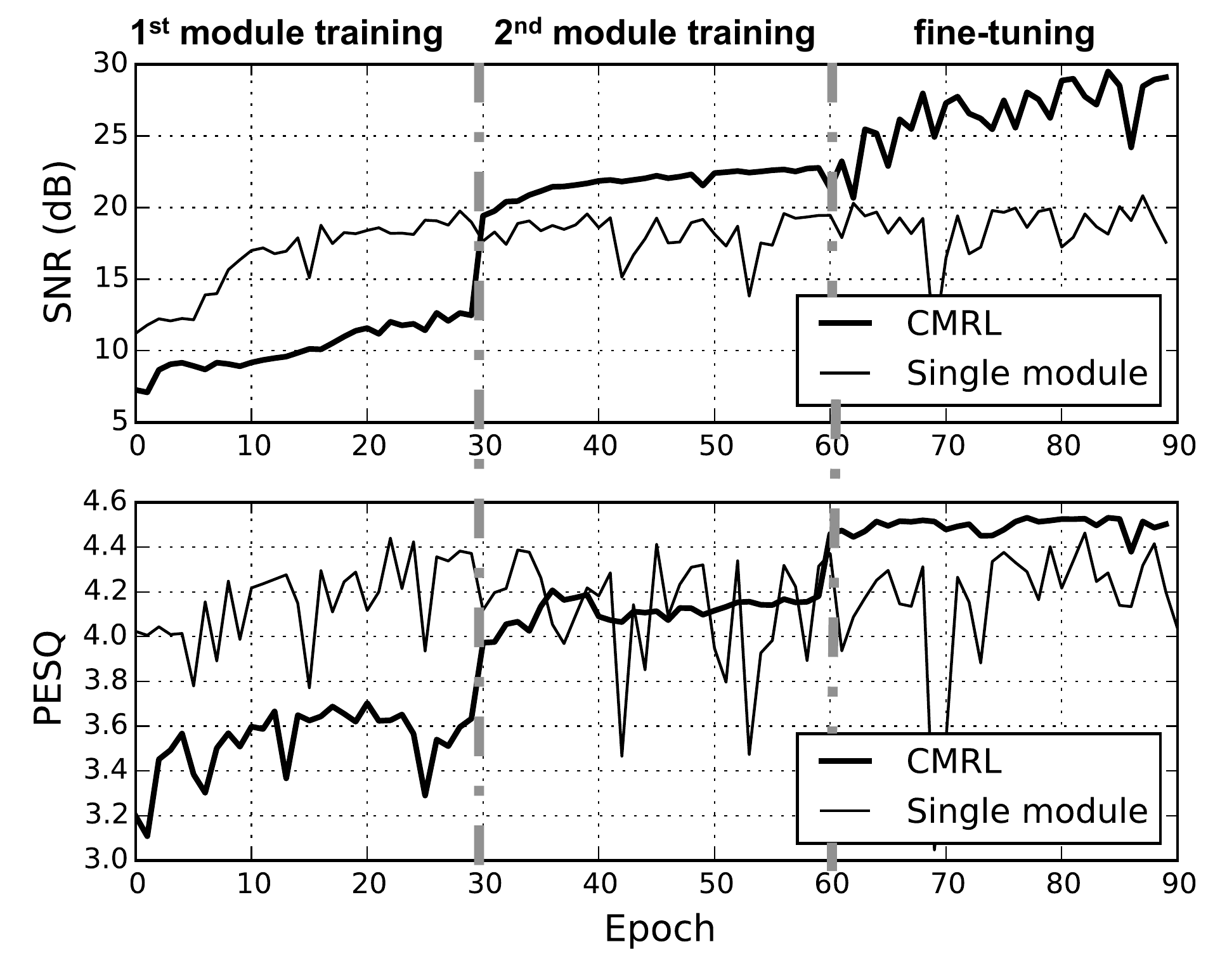}} 
   \subfigure[]{\includegraphics[width=.22\columnwidth]{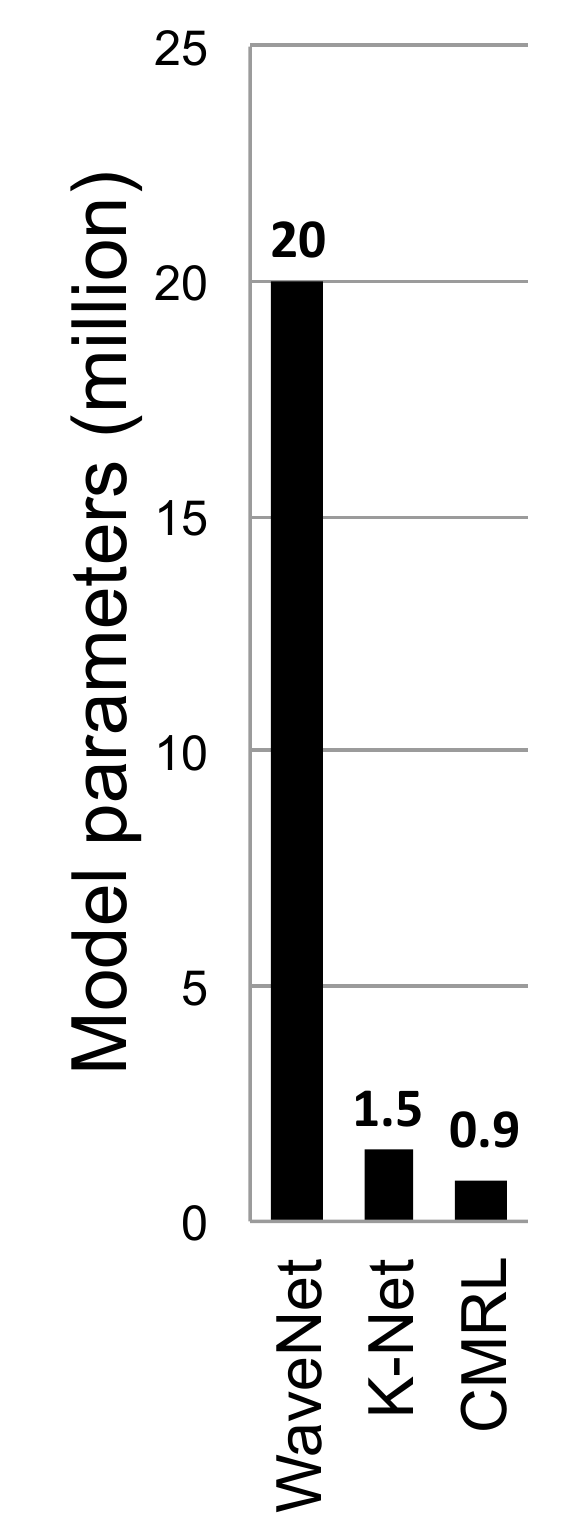}} 
    \caption{(a) SNR and PESQ per epoch (b) model complexity}
    \vspace{-0.06in}
    \label{fig:numparam}
\end{figure}

\subsection{Objective test}

We evaluate 500 decoded utterances in terms of SNR and PESQ with wide band extension (P862.2) \cite{pesq}.
Figure \ref{fig:numparam} (a) shows the effectiveness of CMRL against a system with a single module in terms of  SNR and PESQ values per epoch. 
The single module is with three more bottleneck blocks and twice more codes for a fair comparison. It is trained for 90 epochs with other hyperparameters are unaltered. For both SNR and PESQ, the plot shows a noticeable performance jump as the second module is included, followed by another jump by finetuning.

Table \ref{tab:snrpesq} compares CMRL with AMR-WB and Kankanahalli-Net at four bitrates for the raw PCM input case. CMRL achieves both higher SNR and PESQ at all four bitrate cases. Note that the SNR for CMRL at 8.85 kbps is greater than AMR-WB at 23.85 kbps. CMRL also gives a better PESQ score at 15.85 kbps than AMR-WB at 23.85 kbps. 

\begin{table}
\caption{SNR and PESQ scores on raw PCM test signals.}
\centering
\small
\setlength\tabcolsep{1.0pt}
\resizebox{1\columnwidth}{!}{
\begin{tabular}{ |c|c|c|c|c|c|c|c|c|c|c|c|c|c| }
 \hline
  {Metrics} &\multicolumn{4}{c|}{{SNR (dB)}}& \multicolumn{4}{c|}{{PESQ} }\\
  \hline
  {Bitrate (kbps)} &{8.85}&{15.85}&{19.85}&{23.85}&{8.85}&{15.85}&{19.85}&{23.85}\\
\hline
{AMR-WB}   &9.82&11.93&12.46&12.73&
3.41&3.99&4.09&4.13\\
{K-Net}    &-&-&-&-&3.63&4.13&4.22&4.30\\
{CMRL}    &\textbf{13.45}&\textbf{16.35}&\textbf{17.18}&\textbf{17.33}&\textbf{3.69}&\textbf{4.21}&\textbf{4.34}&\textbf{4.42}\\\hline
\end{tabular}}
\vspace{-0.05in}
\label{tab:snrpesq}
\end{table}

\subsection{Subject test}

Figure \ref{fig:mos} shows MUSHRA test results done by six audio experts on 10 decoded test samples randomly selected with gender equity. At 19.85 kbps and 23.85 kbps, CMRL with LPC residual inputs outperforms AMR-WB. At lower bitrates though, AMR-WB starts to work better. CMRL on raw PCM is found less favored by listeners. We also compare CMRL with OPUS in the high bitrate where OPUS is known to perform well, and find that CMRL slightly outperforms OPUS\footnote{Samples are available at 
\url{http://saige.sice.indiana.edu/research-projects/neural-audio-coding}}.

\subsection{Model complexity}
 The cross-module residual learning simplifies the topology of each component module. Hence, CMRL has less than $5\%$ of the model parameters compared to the WaveNet based codec \cite{google_wavenet_2019}, and outperforms Kankanahalli-Net with $40\%$ less model parameters. Figure \ref{fig:numparam} (b) summarizes the comparison.

\section{Conclusion}
In this work, we demonstrated that CMRL as a lightweight model carrier for DNN based speech codecs can compete with the industrial standards. By cascading two end-to-end modules, CMRL achieved a higher PESQ score at 15.85 kbps than AMR-WB at 23.85 kbps. We also showed that CMRL can consistently outperform a state-of-the-art DNN codec in terms of PESQ. CMRL is compatible with LPC, by having it as the first pre-processing module and by using its residual signals as the input. CMRL, coupled with LPC, outperformed AMR-WB in 19.85 kbps and 23.85 kbps, and worked better than OPUS at 23.85 kbps in the MUSHRA test.  More work is required to examine other module structures to further improve the performance at low bitrates.
\section{Acknowledgements}
This work was supported by Institute for Information \& communications Technology Promotion (IITP) grant funded by the Korea government (MSIT) (2017-0-00072, Development of Audio/Video Coding and Light Field Media Fundamental Technologies for Ultra Realistic Tera-media). The authors also appreciate Srihari Kankanahalli for valuable discussion and sharing audio samples.

\bibliographystyle{IEEEtran}

\bibliography{mybib}

\begin{thebibliography}{10}
\providecommand{\url}[1]{#1}
\csname url@samestyle\endcsname
\providecommand{\newblock}{\relax}
\providecommand{\bibinfo}[2]{#2}
\providecommand{\BIBentrySTDinterwordspacing}{\spaceskip=0pt\relax}
\providecommand{\BIBentryALTinterwordstretchfactor}{4}
\providecommand{\BIBentryALTinterwordspacing}{\spaceskip=\fontdimen2\font plus
\BIBentryALTinterwordstretchfactor\fontdimen3\font minus
  \fontdimen4\font\relax}
\providecommand{\BIBforeignlanguage}[2]{{%
\expandafter\ifx\csname l@#1\endcsname\relax
\typeout{** WARNING: IEEEtran.bst: No hyphenation pattern has been}%
\typeout{** loaded for the language `#1'. Using the pattern for}%
\typeout{** the default language instead.}%
\else
\language=\csname l@#1\endcsname
\fi
#2}}
\providecommand{\BIBdecl}{\relax}
\BIBdecl

\bibitem{classic_book_audiocoding_mpeg}
P.~Noll, ``{MPEG} digital audio coding,'' \emph{IEEE signal processing
  magazine}, vol.~14, no.~5, pp. 59--81, 1997.

\bibitem{classic_algo_audiocoding_mpeg1}
K.~Brandenburg and G.~Stoll, ``{ISO}/{MPEG}-1 audio: A generic standard for
  coding of high-quality digital audio,'' \emph{Journal of the Audio
  Engineering Society}, vol.~42, no.~10, pp. 780--792, 1994.

\bibitem{classic_book_audiocoding}
K.~R. Rao and J.~J. Hwang, \emph{Techniques and standards for image, video, and
  audio coding}.\hskip 1em plus 0.5em minus 0.4em\relax Prentice Hall New
  Jersey, 1996, vol.~70.

\bibitem{lpc}
D.~O'Shaughnessy, ``Linear predictive coding,'' \emph{IEEE potentials}, vol.~7,
  no.~1, pp. 29--32, 1988.

\bibitem{atal1970adaptive}
B.~S. Atal and M.~R. Schroeder, ``Adaptive predictive coding of speech
  signals,'' \emph{Bell System Technical Journal}, vol.~49, no.~8, pp.
  1973--1986, 1970.

\bibitem{schroeder1985code}
M.~Schroeder and B.~Atal, ``Code-excited linear prediction ({CELP}):
  High-quality speech at very low bit rates,'' in \emph{Acoustics, Speech, and
  Signal Processing, IEEE International Conference on ICASSP'85.},
  vol.~10.\hskip 1em plus 0.5em minus 0.4em\relax IEEE, 1985, pp. 937--940.

\bibitem{amrwb}
B.~Bessette, R.~Salami, R.~Lefebvre, M.~Jelinek, J.~Rotola-Pukkila, J.~Vainio,
  H.~Mikkola, and K.~Jarvinen, ``The adaptive multirate wideband speech codec
  (amr-wb),'' \emph{IEEE transactions on speech and audio processing}, vol.~10,
  no.~8, pp. 620--636, 2002.

\bibitem{opus}
J.-M. Valin, G.~Maxwell, T.~B. Terriberry, and K.~Vos, ``High-quality,
  low-delay music coding in the opus codec,'' \emph{arXiv preprint
  arXiv:1602.04845}, 2016.

\bibitem{makhoul1985vector}
J.~Makhoul, S.~Roucos, and H.~Gish, ``Vector quantization in speech coding,''
  \emph{Proceedings of the IEEE}, vol.~73, no.~11, pp. 1551--1588, 1985.

\bibitem{KimMJ2015icmlw}
M.~Kim and P.~Smaragdis, ``Bitwise neural networks,'' in \emph{International
  Conference on Machine Learning (ICML) Workshop on Resource-Efficient Machine
  Learning}, Jul 2015.

\bibitem{deng2010binary}
L.~Deng, M.~Seltzer, D.~Yu, A.~Acero, A.~Mohamed, and G.~Hinton, ``Binary
  coding of speech spectrograms using a deep auto-encoder,'' in \emph{Eleventh
  Annual Conference of the International Speech Communication Association},
  2010.

\bibitem{new_paper_nn_lowbitrate_2016}
M.~Cernak, A.~Lazaridis, A.~Asaei, and P.~Garner, ``Composition of deep and
  spiking neural networks for very low bit rate speech coding,'' \emph{arXiv
  preprint arXiv:1604.04383}, 2016.

\bibitem{new_paper_google_wavenet_2017}
W.~B. Kleijn, F.~S. Lim, A.~Luebs, J.~Skoglund, F.~Stimberg, Q.~Wang, and T.~C.
  Walters, ``Wavenet based low rate speech coding,'' \emph{arXiv preprint
  arXiv:1712.01120}, 2017.

\bibitem{new_paper_bloombergEndtoEnd}
S.~Kankanahalli, ``End-to-end optimized speech coding with deep neural
  networks,'' \emph{arXiv preprint arXiv:1710.09064}, 2017.

\bibitem{liu2018wavenet}
L.-J. Liu, Z.-H. Ling, Y.~Jiang, M.~Zhou, and L.-R. Dai, ``Wavenet vocoder with
  limited training data for voice conversion,'' in \emph{Proc. Interspeech},
  2018, pp. 1983--1987.

\bibitem{google_wavenet_2019}
Y.~L. Cristina~Garbacea, Aaron van den~Oord, ``Low bit-rate speech coding with
  vq-vae and a wavenet decoder,'' in \emph{Proc. ICASSP}, 2019.

\bibitem{salami1997itu}
R.~Salami, C.~Laflamme, B.~Bessette, and J.-P.~Adoul, ``ITU-TG.  729  annex  a:  reduced  complexity  8  kb/s  CS-ACELP  codecfor digital simultaneous voice and data,'' IEEE Communications Magazine, vol. 35, no. 9, pp. 56–63, 1997.

\bibitem{recommendation2003722}
G.~Recommendation, ``722.2:“wideband coding of speech at around 16 kbit/s
  using adaptive multi-rate wideband (amr-wb)”,'' 2003.

\bibitem{neuendorf2012mpeg}
M.~Neuendorf, M.~Multrus, N.~Rettelbach, G.~Fuchs, J.~Robilliard, J.~Lecomte,
  S.~Wilde, S.~Bayer, S.~Disch, C.~Helmrich \emph{et~al.}, ``{MPEG} unified
  speech and audio coding-the iso/mpeg standard for high-efficiency audio
  coding of all content types,'' in \emph{Audio Engineering Society Convention
  132}.\hskip 1em plus 0.5em minus 0.4em\relax Audio Engineering Society, 2012.

\bibitem{samplernn}
S.~Mehri, K.~Kumar, I.~Gulrajani, R.~Kumar, S.~Jain, J.~Sotelo, A.~Courville,
  and Y.~Bengio, ``Samplernn: An unconditional end-to-end neural audio
  generation model,'' \emph{arXiv preprint arXiv:1612.07837}, 2016.

\bibitem{rnn_codec_2019}
Y.~L. Cristina~Garbacea, Aaron van den~Oord, ``High-quality speech coding with
  samplernn,'' in \emph{Proc. ICASSP}, 2019.

\bibitem{johnston2018improved}
N.~Johnston, D.~Vincent, D.~Minnen, M.~Covell, S.~Singh, T.~Chinen,
  S.~Jin~Hwang, J.~Shor, and G.~Toderici, ``Improved lossy image compression
  with priming and spatially adaptive bit rates for recurrent networks,'' in
  \emph{Proceedings of the IEEE Conference on Computer Vision and Pattern
  Recognition}, 2018, pp. 4385--4393.

\bibitem{schuller2001lossless}
G.~Schuller, B.~Yu, and D.~Huang, ``Lossless coding of audio signals using
  cascaded prediction,'' in \emph{2001 IEEE International Conference on
  Acoustics, Speech, and Signal Processing. Proceedings (Cat. No. 01CH37221)},
  vol.~5.\hskip 1em plus 0.5em minus 0.4em\relax IEEE, 2001, pp. 3273--3276.

\bibitem{he2016deep}
K.~He, X.~Zhang, S.~Ren, and J.~Sun, ``Deep residual learning for image
  recognition,'' in \emph{Proceedings of the IEEE conference on computer vision
  and pattern recognition}, 2016, pp. 770--778.

\bibitem{agustsson2017soft}
E.~Agustsson, F.~Mentzer, M.~Tschannen, L.~Cavigelli, R.~Timofte, L.~Benini,
  and L.~V. Gool, ``Soft-to-hard vector quantization for end-to-end learning
  compressible representations,'' in \emph{Advances in Neural Information
  Processing Systems}, 2017, pp. 1141--1151.

\bibitem{shi2016real}
W.~Shi, J.~Caballero, F.~Husz{\'a}r, J.~Totz, A.~P. Aitken, R.~Bishop,
  D.~Rueckert, and Z.~Wang, ``Real-time single image and video super-resolution
  using an efficient sub-pixel convolutional neural network,'' in
  \emph{Proceedings of the IEEE conference on computer vision and pattern
  recognition}, 2016, pp. 1874--1883.

\bibitem{tan2018gated}
K.~Tan, J.~Chen, and D.~Wang, ``Gated residual networks with dilated
  convolutions for supervised speech separation,'' in \emph{Proc. ICASSP},
  2018.

\bibitem{huffman1952method}
D.~A. Huffman, ``A method for the construction of minimum-redundancy codes,''
  \emph{Proceedings of the IRE}, vol.~40, no.~9, pp. 1098--1101, 1952.

\bibitem{witten1987arithmetic}
I.~H. Witten, R.~M. Neal, and J.~G. Cleary, ``Arithmetic coding for data
  compression,'' \emph{Communications of the ACM}, vol.~30, no.~6, pp.
  520--541, 1987.

\bibitem{welch1986error}
L.~R. Welch and E.~R. Berlekamp, ``Error correction for algebraic block
  codes,'' Dec.~30 1986, uS Patent 4,633,470.

\bibitem{itu20031534}
R.~B. ITU-R, ``1534-1,“method for the subjective assessment of intermediate
  quality levels of coding systems (mushra)”,'' \emph{International
  Telecommunication Union}, 2003.

\bibitem{timit}
J.~S. Garofolo, L.~F. Lamel, W.~M. Fisher, J.~G. Fiscus, D.~S. Pallett, N.~L.
  Dahlgren, and V.~Zue, ``{TIMIT} acoustic-phonetic continuous speech corpus,''
  \emph{Linguistic Data Consortium, Philadelphia}, 1993.

\bibitem{kingma2014adam}
D.~P. Kingma and J.~Ba, ``Adam: A method for stochastic optimization,''
  \emph{arXiv preprint arXiv:1412.6980}, 2014.

\bibitem{pesq}
A.~Rix, J.~Beerends, M.~Hollier, and A.~Hekstra, ``Perceptual evaluation of
  speech quality ({PESQ})-a new method for speech quality assessment of
  telephone networks and codecs,'' in \emph{Acoustics, Speech, and Signal
  Processing, 2001. Proceedings.(ICASSP'01). 2001 IEEE International Conference
  on}, vol.~2.\hskip 1em plus 0.5em minus 0.4em\relax IEEE, 2001, pp. 749--752.

\end{thebibliography}

\end{document}